\documentclass[preprint,authoryear,12pt]{elsarticle}

\usepackage{amssymb}

\journal{Astronomy and Computing}

\begin{document}

\begin{frontmatter}



\title{ASERA: A Spectrum Eye Recognition Assistant for Quasar Spectra}


\author[label1]{Hailong YUAN}
\author[label1]{Haotong ZHANG}
\author[label1]{Yanxia ZHANG}
\author[label1]{Yajuan LEI}
\author[label1]{Yiqiao DONG}
\author[label1]{Yongheng ZHAO}

\address[label1]{Key Laboratory of Optical Astronomy, National Astronomical Observatories, Chinese Academy of Sciences, 20A Datun Road, Chaoyang District, Beijing, China, 100012. Email: yuanhl@bao.ac.cn}

\begin{abstract}
Spectral type recognition is an important and
fundamental step of large sky survey projects in the data reduction
for further scientific research, like parameter measurement
and statistic work. It tends out to be a huge job to manually
inspect the low quality spectra produced from massive spectroscopic
survey, where the automatic pipeline may not provide confident
type classification results. In order to improve the efficiency and
effectiveness of spectral classification, we develop a
semi-automated toolkit named ASERA, \emph{A} \emph{S}pectrum \emph{E}ye \emph{R}ecognition \emph{A}ssistant.
The main purpose of ASERA is to help the user in quasar spectral recognition and redshift measurement.
Furthermore it can also be used to recognize various types of spectra of stars, galaxies and AGNs (Active Galactic Nucleus).
It is an interactive software allowing the user to visualize observed spectra,
superimpose template spectra from the Sloan Digital Sky Survey (SDSS),
and interactively access related spectral line information.
It is an efficient and user-friendly toolkit for accurate classification of spectra observed by LAMOST (the Large Sky Area Multi-object Fiber Spectroscopic
Telescope).
The toolkit is available in two modes: a Java standalone application and a Java applet.
ASERA has a few functions, such as wavelength and flux scale setting, zoom in and out, redshift estimation, spectral line identification,
which helps user to improve the spectral classification accuracy especially for low quality spectra and reduce the labor of eyeball check.
The function and performance of this tool is displayed through the recognition of several quasar spectra and a late type stellar spectrum from the LAMOST Pilot survey.
Its future expansion capabilities are discussed.
\end{abstract}


\end{frontmatter}


\section{Introduction}
\label{Introduction}
The Large Sky Area Multi-object Fiber Spectroscopic Telescope
(LAMOST) is a special reflecting Schmidt telescope specialized for
conducting spectroscopic surveys with a wide field of view and a
large aperture \citep{Wang:96}. One of the key scientific goals of
LAMOST is the extragalactic spectroscopic survey of the large scale
structure of the Universe and the physics of galaxies and quasars
\citep{2011AAS...21812307W}. The pilot survey \citep{2012RAA....12.1197C} performed from
October 2011 to June 2012 and the regular survey started in
September 2012. There have been already millions of targets
observed, including thousands of quasar candidates. Then to recognize quasars via spectra becomes essential for
critical candidate confirmation and follow up scientific work.

Spectra with high signal-to-noise ratio (SNR) are easily classified and the physical parameters can be determined with high accuracy using the LAMOST data processing pipeline \citep{2012RAA....12.1243L}.
However there are still a large number of spectra with low SNR and probably some defects (e.g., skylight residual, splice connecting red part and blue part).
Before the automatic pipeline is upgraded to be intelligent enough, eyeball check is in great need and enough astronomical knowledge is necessary.
In SDSS quasar survey, visual inspection has been widely used to ensure the reliability of spectral identifications (P$\hat{a}$ris et~al. 2012).
However in a large sky spectra survey, the quantity of spectra is very large.
In order to reduce human efforts,
we developed a spectrum eye recognition toolkit which provides a flexible platform to help identifying quasar spectra and estimating their redshifts at the same time.
We call this toolkit ASERA.
Both fits-formatted and image-formatted spectral files are supported.
The input source can be placed in a local storage device, or distributed on the internet, described by a URL name.
Since the software is developed using the Java programming language, it can be either started as a desktop application or accessed via a web browser,
after deploying it as a Java applet.
ASERA is initially dedicated to experienced spectrum analysts.
It can also be used by teachers, undergraduate students and amateur astronomers.

In the following sections, we first describe the detailed design and realization of the toolkit.
Then several examples and figures are presented to explain how to use this toolkit on the spectra from LAMOST.
In the end we discuss the system error on the estimated redshift $z$ and the following upgrade plans to extend the functionality of ASERA.

\section{ASERA Development Status}
The basic design idea of this toolkit is to overlay the quasar spectral template on the observed spectrum.
With the help of the spectral identification experience, researchers try to superimpose these two spectra by choosing appropriate redshift and flux scale ratio.
The first important step is to choose a quasar template.

\subsection{The quasar spectrum template}\label{subsection_quasartemp}
\begin{figure}[h]
\begin{center}
\includegraphics[width=0.5\textwidth]{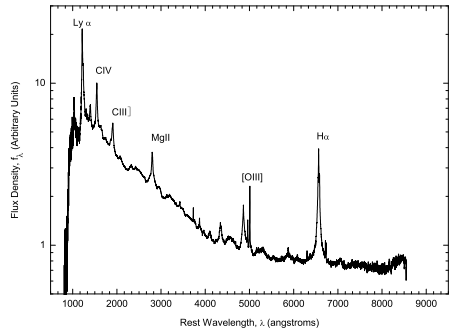}
\caption{\small The median composite quasar spectrum template from SDSS \citep{2001AJ....122..549V}.}
\label{fig:qsotemp}
\end{center}
\end{figure}

Due to the similar wavelength coverage and spectral resolution of LAMOST and SDSS,
a median composite quasar spectrum generated by a sample of over 2200 quasars from SDSS is applied as the standard quasar template \citep{2001AJ....122..549V}.
The spectrum, as shown in Figure~\ref{fig:qsotemp}, has over 80 identified emission lines within the band of wavelength from about 900 to 9000 angstroms.
Eight most distinct emission lines are listed in Table 1.

\begin{table}[h]
\begin{center}
\caption{\small Eight most distinct emission lines of the composite quasar (Table 2 of \citet{2001AJ....122..549V}).}
\label{qsolines}
\vspace{0.5cm}
\begin{tabular}{lcc}
\hline ID & $\lambda_{obs}$(\r{A}) \\
\hline
Ly$\alpha$ & 1216.25$\pm$0.37 \\
C IV       & 1546.15$\pm$0.14 \\
C III      & 1905.97$\pm$0.12 \\
Mg II      & 2800.26$\pm$0.10 \\
H$\beta$   & 4862.66* \\
O III      & 4960.36$\pm$0.22 \\
O III      & 5008.22$\pm$0.17 \\
H$\alpha$  & 6564.93$\pm$0.22 \\
\hline
\end{tabular}
\medskip\\
\small
$*$The wavelength of H$\beta$ is from Table 4 of \citet{2001AJ....122..549V}.
\end{center}
\end{table}

In this toolkit, the composite quasar spectrum can be transformed to
various shapes by adjusting the redshift and the flux scale via the mouse.
By comparing the observed spectrum with the transformed composite spectrum,
researchers can easily figure out the best fit and provide the
apt redshift value if the observed target is a quasar.

\subsection{Redshift calculation}
\label{p2.2}
Since the spectrum point is described by wavelength and flux
density, a simple linear transformation is imported to get the pixel
coordinate, and inversely to get spectrum point from pixel
coordinate. We construct a linear pixel transformation function from the
observed target spectrum firstly and then apply it to the template.

In our toolkit, a default pixel transformation function is applied for the LAMOST spectrum products.
To use this toolkit for image spectra from other spectral surveys, the user can choose
the starting and end points in the pixel frame by simple mouse
click, and tag them with the correct wavelength value. Then the new
transformation can be established on these two points.

The first step for plotting the template spectrum is to define the redshift $z$.
We will choose one of the emission lines in Table \ref{qsolines} as the \textbf{reference line} and place it on the observed target spectrum.
The redshift $z$ of a wavelength which is shifted from $\lambda_0$ (from Table~\ref{qsolines}) to $\lambda$ (calculated using pixel transformation function inversely) is defined as
\begin{equation}
z=\frac{\lambda-\lambda_0}{\lambda_0}
\label{eq1}
\end{equation}
On the contrary, given the redshift value $z$, the wavelength $\lambda_0$
of a composite quasar spectrum can be shifted to $\lambda$ defined
by
\begin{equation}
\lambda=\lambda_0\times(z+1)
\label{eq2}
\end{equation}
Then the whole template spectrum can be plotted at the specified redshift and flux scale.
The user can find the most possible spectral type of the target and its corresponding redshift by changing and adjusting the template spectrum interactively.

\subsection{Supported spectral formats and locations}
The current version of this toolkit supports both image-formatted and fits-formatted spectral files produced by LAMOST.
File types are identified by their suffixes.

Files ending with ``PNG", ``JPG", ``JPEG", ``WBMP", ``GIF" and ``BMP" are treated as images.
The image file is plotted directly on a plane as the background and then the template spectrum as the foreground.
The pixel transformation mentioned in the subsection \ref{p2.2} is used to calculate wavelength of any pixel point.

Files ending with ``FITS" or ``FIT" are treated as LAMOST fits products.
The java fits library ``nom.tam.fits" is imported to provide I/O for FITS image and binary tables.
Then the spectrum is obtained by referring to the fits header definition of the LAMOST spectra.
In the latest version, ``fits" files from SDSS are also supported.

The data file can be placed in a local storage device or on the internet.
The Uniform Resource Locator (URL) is used to describe both the exact location and accessing protocol.
The standard edition of Java Develop Kit supports several protocol types, such as ``file", ``http", ``ftp" and ``gopher".
The ``file", ``ftp" and ``http" protocol type have been tested for the current version.
Here are some examples of the URL external format string used in Windows platform:\\
\emph{
\tiny
ftp://user:pwd@data.lamost.org/pdr/fits/20111024/F5902/spec-55859-F5902\_sp16-249.fits\\
http://data.lamost.org/pdr/fits/20111024/F5902/spec-55859-F5902\_sp16-249.fits\\
file://F:$\backslash$B55878$\backslash$spec-55878-B87808\_sp02-002.fits
}

\subsection{Installation}
The local installation of the toolkit takes merely no time since the latest version is provided as a single JAR file.
Java Virtual Machine (JVM) is required to run the toolkit.
Once the JVM is installed, the toolkit can be easily launched on all the dominating operation systems such as Windows, Linux and Mac-OS.

Users can also start the toolkit as an applet from a simple web navigator such as Internet Explorer and Firefox.
Any other software component can extend the capability of this toolkit by an HTTP link to the server.
To enable this ability, the program should be firstly deployed as a Java applet in a dynamic web server, such as Apache Tomcat Server and Apache HTTP Server.
The benefit for this deployment type is to save time for the client side user since the web server manager need handle the upgrade of the software release.
However the weakness is the dependence of net access to the web server.
Here is the example code for embedding the applet in a Java Server Page (JSP) file, namely ``index.jsp":

\emph{
$<$applet
code=``FittingApplet.class"
archive=``ASERA.jar"
width=``900" height=``720"
$/>$
}

In this condition, the JAR library file should be placed in the same directory as the JSP file.

\subsection{The graphical interface}
\begin{figure}[h]
\begin{center}
\includegraphics[width=0.5\textwidth]{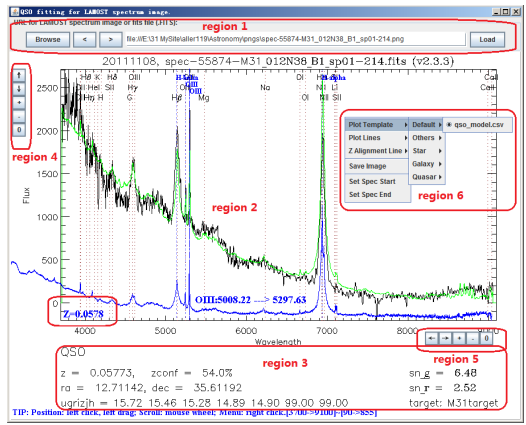}
\caption{\small The graphical interface of ASERA: the main window which consists of six regions.}\label{fig:interface2}
\end{center}
\end{figure}

ASERA offers a single main window for displaying and manipulating the data, as shown in Figure~\ref{fig:interface2}.
The combined functional regions are discussed in the following paragraphs.

Region 1: the input data resource path and the path specification buttons.

Region 2: the target and template spectrum in the centric viewport.
The final redshift $z$ of the template spectrum is printed at the lower-left corner.
A set of spectral lines are plotted together with the template spectrum at the same time.

Region 3: the most important information of the target including RA, DEC, target name, target type, SNR, magnitude, et al.

Region 4: a button group for scaling and shifting flux range of the viewport for the target spectrum.
It helps the user to acquire a proper flux density range for inspection.

Region 5: a button group for scaling and shifting wavelength range of the viewport for both the target spectrum and the template spectrum.
It helps the user to acquire a proper wavelength range for inspection.

Region 6: the mouse right click popup menu.
It provides a set of functions including template spectrum selection, visible absorption/emission lines selection, redshift \textbf{reference line} selection, image saving, pixel-wavelength conversion starting and end point specification.
The current pixel-wavelength conversion status is displayed in the bottom line of the main window.

The adjustment of the flux density of the template spectrum is handled by mouse.
A left mouse button click event will replace the flux zero point and the \textbf{reference line} of the template to the clicked position.
Then the redshift $z$ will be recalculated and the spectrum will be repainted.
A left mouse button drag event will shift the flux zero point and the \textbf{reference line} according to the drag distance.
The mouse scrolling event will change the flux scale.

\begin{figure}[h]
\begin{center}
\includegraphics[width=0.5\textwidth]{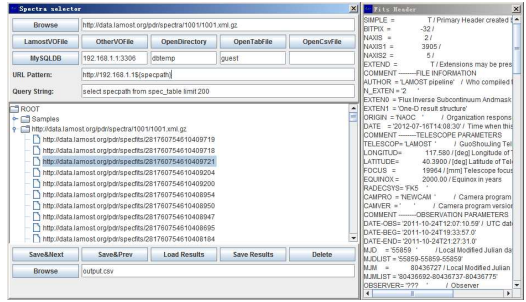}
\caption{\small
The spectrum selector and the FITS header viewer of ASERA.
Currently the pilot survey spectra of LAMOST are available at ``http://data.lamost.org/sas/pdr/spectra/".
}
\label{fig:interface3}
\end{center}
\end{figure}

In addition, the toolkit provides a spectrum selector and a FITS header viewer component, as shown in Figure~\ref{fig:interface3}.
The spectrum selector enables the user to open a batch of spectra at one time and presents them in a tree like component.
Currently there are three approaches to generate a spectral file list.
The first is to scan the local directory containing fits and image files.
The second is to parse the textual or XML VOTable file containing spectral URLs.
The third is to construct URLs by querying a MySQL Database.
Especially, the URL locating the VOTable can be either a static XML file or a dynamic web service, for example the SSAP (Simple Spectral Access Protocol) \citep{2004SPIE.5493..262D} server.
The common compressed file formats, such as ZIP and GZIP, are recognized.
The FITS header viewer provides a detailed FITS header information for users.

\section{ASERA Application}

\subsection{Quasar spectral recognition}
\begin{figure}[!!!h]
\begin{center}
\includegraphics[width=0.5\textwidth]{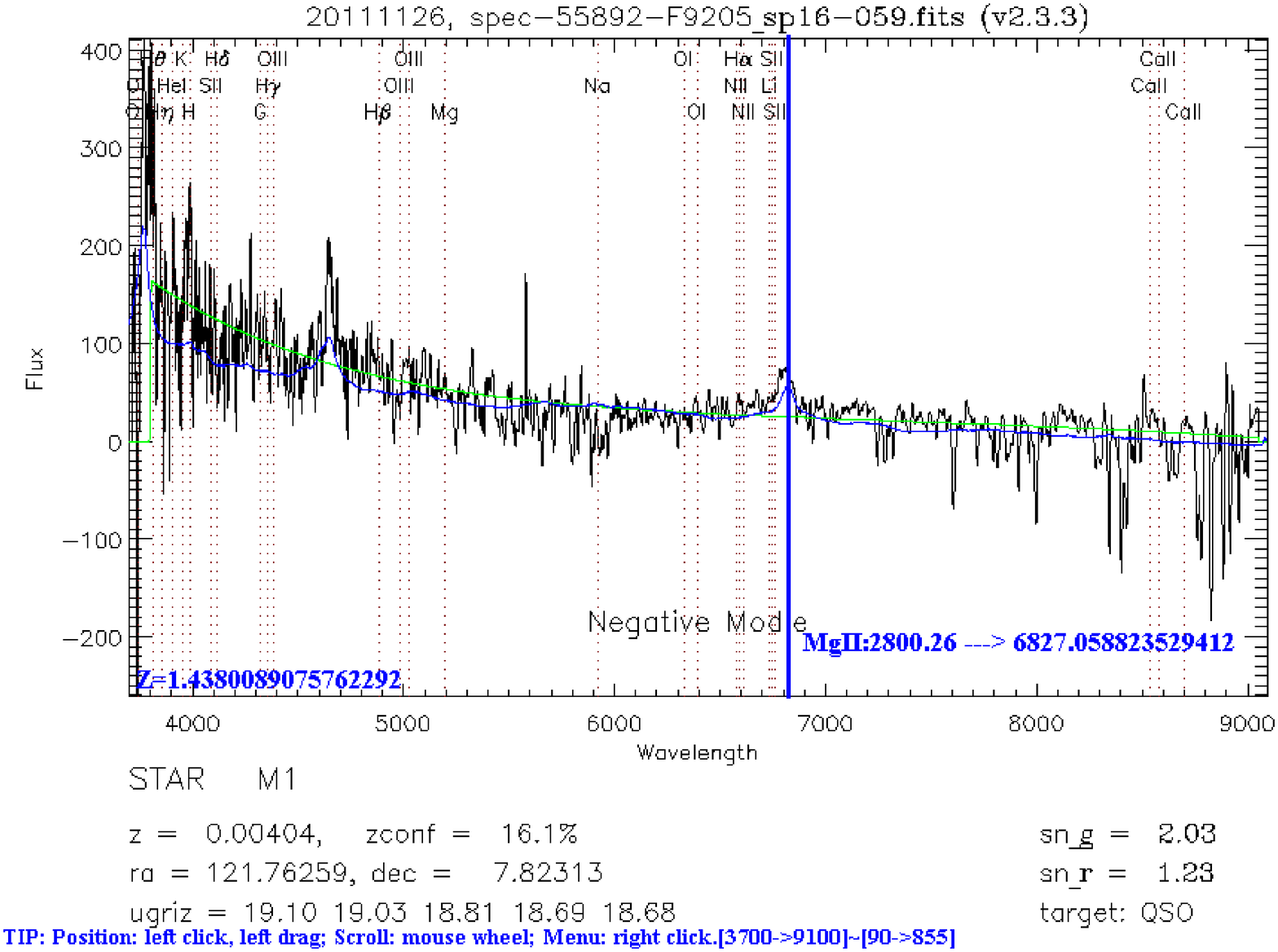}
\includegraphics[width=0.5\textwidth]{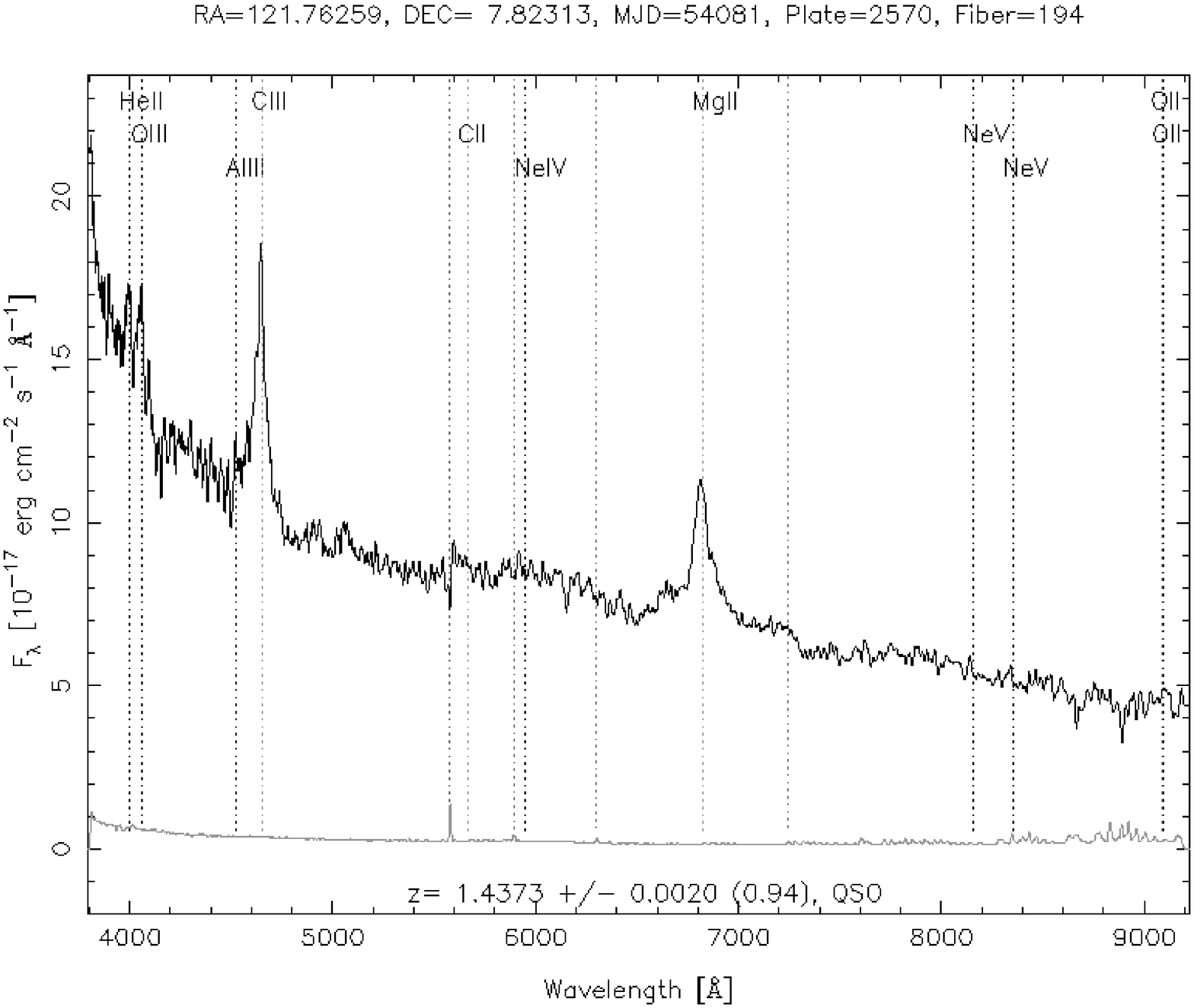}
\caption{\small
The LAMOST spectral identification using the toolkit and comparison with the SDSS spectrum.
The upper panel shows a spectrum of a target observed by LAMOST on 26 November, 2011.
The PLANID is ``F9205", SPECID is 16 and FIBERID is 59.
The SNR of this spectrum is low but the researchers can still recognize the broad emission line.
The input catalog includes this target as a quasar candidate.
Because of the low SNR, the pipeline failed to give a confident classification.
The redshift given by this toolkit is 1.43801.
The lower panel shows the spectrum of the same target given by SDSS DR7, which is available at
``http://cas.sdss.org/dr7/en/tools/quicklook/quickobj.asp?ra=121.76259$\&$dec=7.82313".
The redshift given by SDSS is 1.43726. }\label{fig:result1}
\end{center}
\end{figure}

\begin{figure}[!!!h]
\begin{center}
\includegraphics[width=0.5\textwidth]{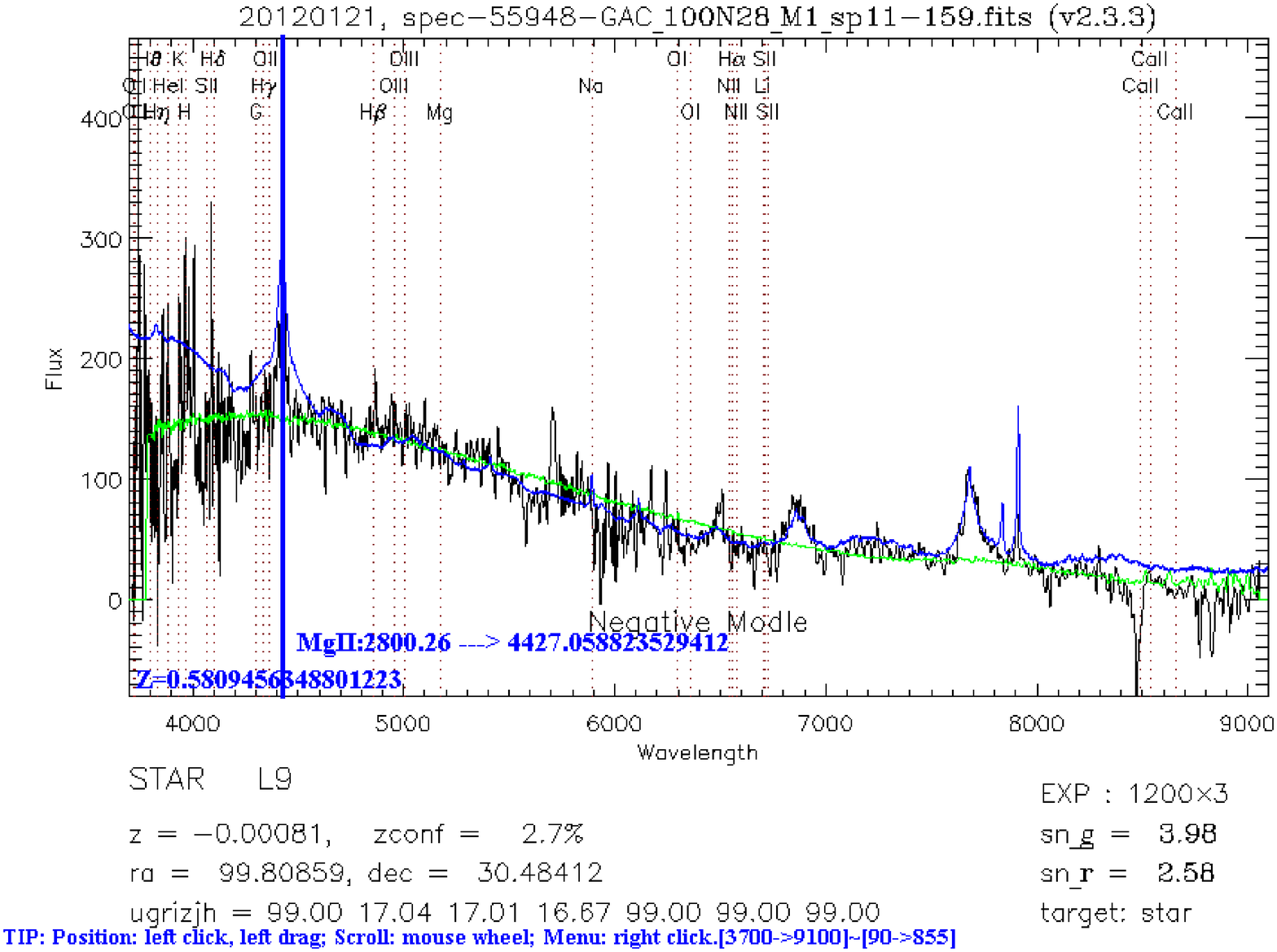}
\includegraphics[width=0.5\textwidth]{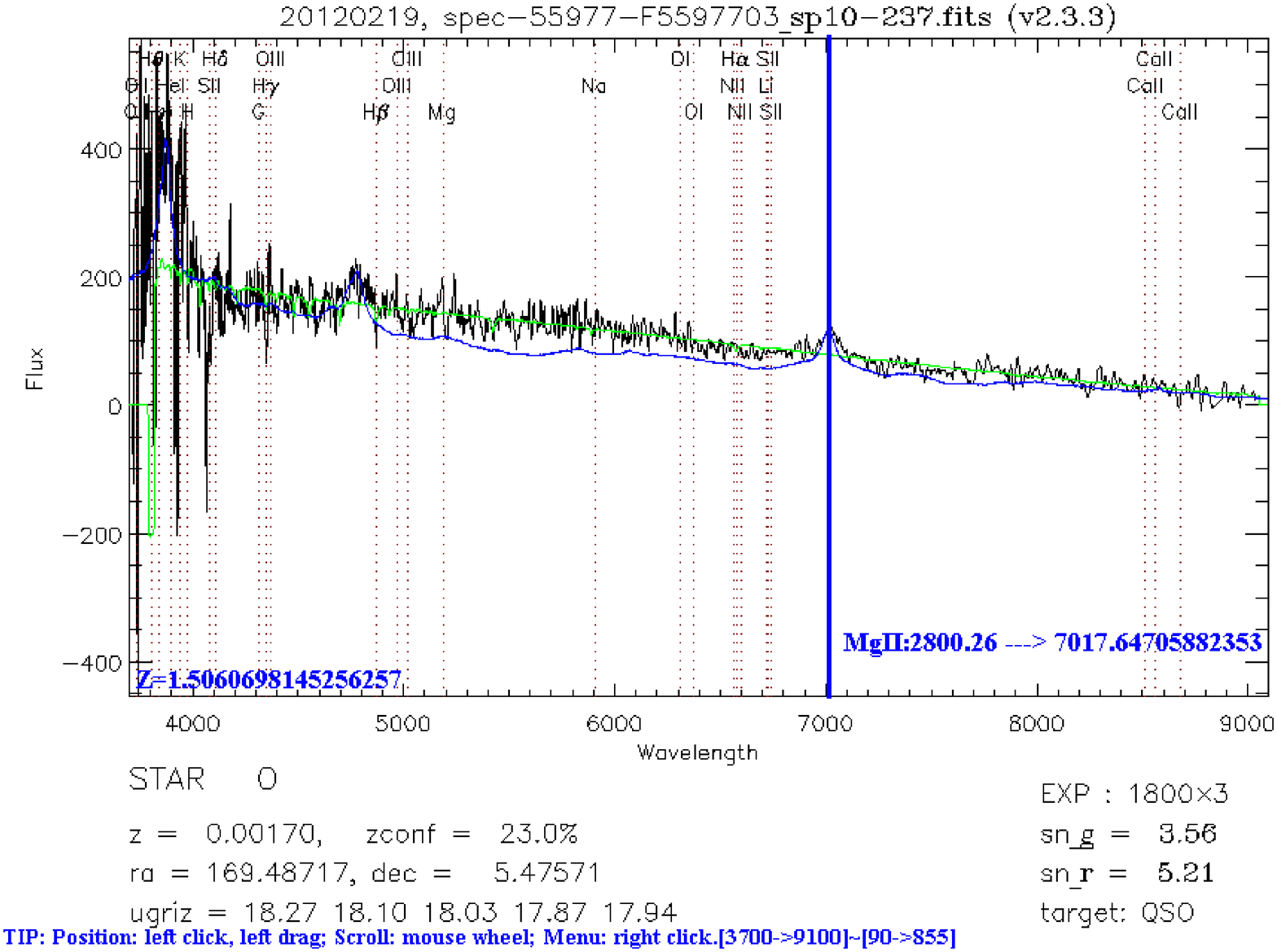}
\caption{\small
Two examples of LAMOST quasar spectral identifications using the toolkit.
The first target was observed on 21 January, 2012.
The PLANID is ``GAC$\_$100N28$\_$M1", SPECID is 11 and FIBERID is 159.
The second target was observed on 19 February, 2012.
The PLANID is ``F5597703", SPECID is 10 and FIBERID is 237.
}
\label{fig:result2}
\end{center}
\end{figure}

In the pilot survey about 400 plates were observed including several thousands of quasar candidates.
SNR of some spectra are very low thus most of the features can't be recognized except the broad emission line.
Some spectra have a little bit higher SNR but suffer from sky emission line residuals.
Automatic program always fail to classify those spectral types and determine the redshifts with high confidence.
With this toolkit, users can simplify the discrimination process under the guide of their rich spectral recognition experience.
In Figure~\ref{fig:result1}, we pick a spectrum processed by an early version pipeline to test our toolkit.
The observed data is shown in black; the green line is the fitting result of the pipeline; the blue line is plotted by this toolkit; the blue vertical line represents the position of the \textbf{MgII} emission line.
Apparently, the spectrum was misclassified as ``star" by the pipeline, but can be identified as ``quasar" with the help of this toolkit.
By means of this toolkit, several emission lines are easily found, meanwhile the redshift can be obtained handily and is printed at the lower-left corner of Figure~\ref{fig:result1}.
Besides, the spectrum of the same source observed by SDSS is presented for comparison.
The spectrum from SDSS is apparently identified as a quasar while the spectrum from LAMOST is difficult to recognize.
To further demonstrate the feasibility of this toolkit, two objects misclassified as ``star" by the pipeline are identified as ``quasar" with the help of this toolkit, as shown in Figure~\ref{fig:result2}.

\subsection{Spectral recognition of other types of objects}

\begin{figure}[h]
\begin{center}
\includegraphics[width=0.5\textwidth]{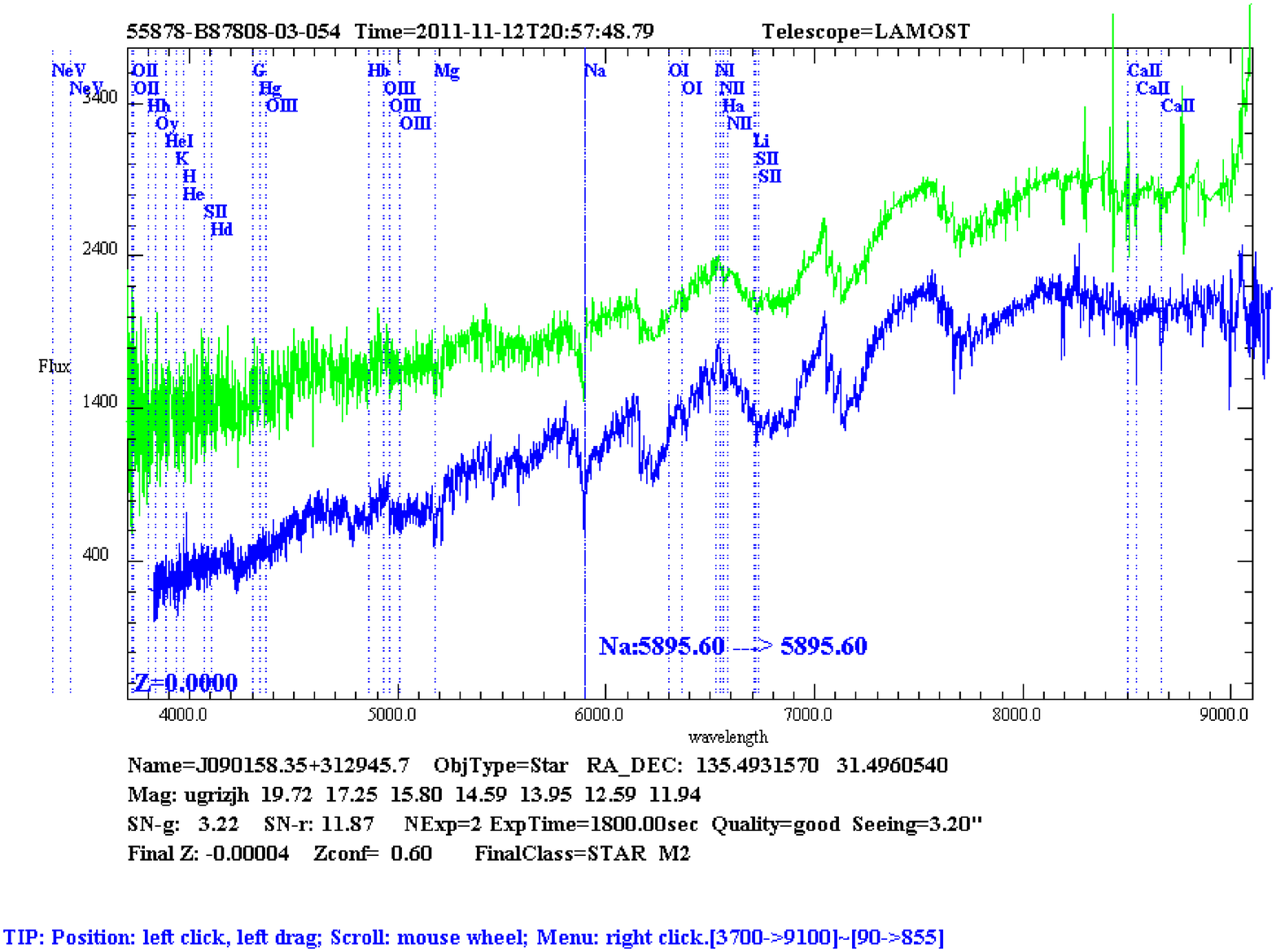}
\caption{\small
The recognition of an M-type star spectrum from LAMOST pilot survey using ASERA.
The upper spectrum is the target spectrum from LAMOST pilot survey.
The PLANID is ``B87808", the SPECID is 3 and the FIBERID is 54.
The lower spectrum is the template of a Late-Type star, M or later, from SDSS, which is available at ``http://www.sdss.org/dr5/algorithms/spectemplates/spDR2-012.fit".
Note these two spectra are plotted with a flux gap by design to
help inspection. }
\label{fig:result3}
\end{center}
\end{figure}

By importing other templates, ASERA can be used to recognize spectra from various types of celestial bodies.
The SDSS has provided 33 typical spectral templates in ``http://www.sdss.org/dr5/algorithms/spectemplates/", including various types of stars, galaxies and quasars.
The template wavelength has already been transformed to rest frame, the same as the composite quasar spectrum mentioned in section~\ref{subsection_quasartemp}.
For templates whose redshifts are not absolutely zero, their wavelength is recalculated when loaded.
In Figure~\ref{fig:result3}, we show an example of ASERA to recognize an M-type star spectrum from the LAMOST pilot survey.
The difference between the recognition of quasars and stars is that the later needs little redshift adjustment.

\section{Discussion}
In this toolkit, the redshift $z$ is calculated from the pixel coordinate, thus the redshift systematic error is
\begin{equation}
Err\_z=({\Delta}N{\times}k)/{\lambda}_0
\label{eq3}
\end{equation}
here
$\lambda_0$ is the wavelength of the emission line that we choose from Table 1 as the \textbf{reference line},
k is the wavelength difference between two adjacent pixels and
$\Delta$N is the difference of pixel between the point we choose and the ideal correct point.
For example when the wavelength varies from 3700 to 9100 and the pixel width is 765, k is about 7.05882.
Taken the lines in Table 1 as examples, with an assumed $\Delta$N of 1, the system errors vary from 0.001075 to 0.005804.

In future, we are ready to update the toolkit in several approaches.
Firstly we will import more spectral templates of other types of celestial objects together with an interface to load a user specified template.
Secondly we will extend the supported spectral data formats from most important survey services, besides LAMOST and SDSS.
The astronomical data is complicated and the format is hard to be unified.
The IVOA has already released many data representation and accessing protocols to facilitate the communication but time is needed for popularization and application.
The VOTable from LAMOST data release server can be recognized currently but we need to extend the access to the online spectral service using the Simple
Spectral Access Protocol (SSAP) proposed by the IVOA.
Spectra in VOTable format will also be recognized and processed.

\section{Conclusions}
To improve the efficiency and effectiveness of spectral classification, ASERA, a spectrum eye recognition assistant, is developed using Java programming language, especially designed for quasar spectral recognition.
The toolkit includes
a graphical interactive interface with both the target spectrum and the template spectrum plotted,
a group of convenient viewport adjustment functions to provide entire or partial inspection of the spectrum arbitrarily,
and various spectral templates helping users to identify the target spectrum by eye.
Via choosing a suitable redshift $z$ interactively, an artificial spectrum can be generated from a composite spectrum from Sloan Digital Sky Survey (SDSS).
By comparing the generated spectrum with the target spectrum, taking the human experience as reference, users can finally recognize whether the target spectrum is a quasar or not, without being hampered by the partial abnormal or low SNR spectra.
At the same time, ASERA may estimate the redshift $z$ of the recognized quasar spectrum.
Several quasar spectra from the LAMOST Pilot survey are tested to show the advantage of this toolkit in handling low SNR spectra with skylight residual or stray light.
ASEAR can be used to recognized various types of stars, galaxies and AGNs by importing their related template spectra.
The systematic error of the redshift calculation is discussed.
The toolkit will be publicly available as soon as possible and user may contact the author for a trial edition at present.
In the future, FITS spectral files besides LAMOST and SDSS, will be supported further.
Also, we will realize the access to the online spectral service using the Simple Spectral Access Protocol (SSAP) proposed by the IVOA.
In addition, spectra in VOTable format will also be recognized and processed.

\section*{Acknowledgments}
This paper is funded by the National Natural Science Foundation of China under grant Nos.10778724, 11178021 and No.11033001. We acknowledge LAMOST and SDSS databases.
Guoshoujing Telescope (the Large Sky Area Multi-Object Fiber Spectroscopic Telescope LAMOST) is a National Major Scientific Project built by the Chinese Academy of Sciences. Funding for the project has been provided by the National Development and Reform Commission. LAMOST is operated and managed by the National Astronomical Observatories, Chinese Academy of Sciences.

\bibliographystyle{model3-num-names}







\end{document}